\documentclass{svproc}

\usepackage{url}
\usepackage{amsmath}
\usepackage{amssymb}
\usepackage{graphicx}
\usepackage{array}
\usepackage{multicol}
\newcommand{\PreserveBackslash}[1]{\let\temp=\\#1\let\\=\temp}
\usepackage{multirow}
\usepackage{booktabs}
\newcolumntype{C}[1]{>{\PreserveBackslash\centering}p{#1}}
\newcolumntype{R}[1]{>{\PreserveBackslash\raggedleft}p{#1}}
\newcolumntype{L}[1]{>{\PreserveBackslash\raggedright}p{#1}}

\begin{document}
\mainmatter              
\title{Disentangled Dual-Branch Graph Learning for Conversational Emotion Recognition}
\titlerunning{Hamiltonian Mechanics}  
%
\author{Chengling Guo\inst{1} \and Yuntao Shou\inst{1}
Tao Meng\inst{1} \and Wei Ai\inst{1} \and Yun Tan\inst{2} \and Keqin Li\inst{3}}
\authorrunning{Chengling Guo et al.} 
%
%
\institute{College of Computer and Mathematics, Central South University of Forestry and Technology \\
\email{shouyuntao@stu.xjtu.edu.cn},
\and
Changsha Hospital for Maternal and Child Health Care 
\and
Department of Computer Science, State University of New York}

\maketitle              

\begin{abstract}  
Multimodal emotion recognition in conversations aims to infer utterance-level emotions by jointly modeling textual, acoustic, and visual cues within context. Despite recent progress, key challenges remain, including redundant cross-modal information, imperfect semantic alignment, and insufficient modeling of high-order speaker interactions. To address these issues, we propose a framework that combines dual-space feature disentanglement with dual-branch graph learning. A shared encoder and modality-specific encoders are used to separate modality-invariant and modality-specific representations. The invariant features are modeled by a Fourier graph neural network to capture global consistency and complementary patterns, with a frequency-domain contrastive objective to enhance discriminability. In parallel, a speaker-aware hypergraph is constructed over modality-specific features to model high-order interactions, along with a speaker-consistency constraint to maintain coherent semantics. Finally, the two branches are fused for utterance-level emotion prediction. Experiments on IEMOCAP and MELD demonstrate that the proposed method achieves superior performance over strong baselines, validating its effectiveness.
\end{abstract}

\section{Introduction}
Despite recent advances, MERC remains challenging for three main reasons. First, multimodal inputs are heterogeneous and often contain redundant or overlapping emotional information, which can introduce noise and hinder discriminative learning. Second, emotional signals across modalities are not always temporally or semantically aligned \cite{zhang2025multimodal,shou2025cilf,shou2026dual}, making cross-modal interaction difficult. Third, conversational emotions depend not only on local context but also on complex multi-speaker interactions, where pairwise modeling is often insufficient to capture high-order dependencies.

Existing MERC methods mainly improve performance through multimodal fusion and contextual modeling. Early approaches rely on attention mechanisms and unified representation learning to integrate multimodal cues, but they typically project all modalities into a shared space without explicitly separating modality-invariant and modality-specific information. This makes them sensitive to noise, redundancy, and missing signals. Graph-based methods further model conversational structure by capturing dependencies between utterances, while recent advances such as frequency-domain graph learning and hypergraph neural networks enhance the modeling of long-range and high-order relationships. However, most of these methods do not explicitly disentangle shared and private information before graph learning.

To address these limitations, we propose a framework that integrates dual-space feature disentanglement with dual-branch graph learning. Specifically, a shared encoder is used to capture modality-invariant emotional representations, while private encoders preserve modality-specific features, enabling effective feature decoupling. Based on this, a dual-branch architecture is constructed, where a Fourier graph neural network models global emotional patterns from invariant features, and a hypergraph neural network captures higher-order speaker interactions from specific features. In addition, a frequency-domain contrastive learning strategy and a speaker consistency constraint are introduced to enhance feature discrimination and maintain semantic consistency across dialogues. Finally, the representations from both branches are fused for emotion classification. The contributions of this study can be summarized as follows:

\begin{itemize}
    \item We propose a dual-space feature disentanglement framework that explicitly separates modality-invariant emotions from modality-specific cues via shared and private encoders, effectively reducing cross-modal redundancy and enhancing robustness.
    
    \item We develop a dual-branch graph learning architecture where a Fourier GNN captures long-range frequency consistency, and a speaker-aware Hypergraph NN models higher-order conversational interactions, addressing both long-distance dependencies and complex contextual dynamics.
    
    \item We introduce a frequency-domain contrastive learning strategy to boost feature discriminability, alongside a speaker perception consistency constraint that ensures prediction coherence for utterances by the same speaker.
    
    \item Extensive experiments on IEMOCAP and MELD datasets demonstrate that our method outperforms state-of-the-art approaches in both accuracy and robustness, validating the effectiveness of the proposed framework.
\end{itemize}

\section{Related Work}
\subsection{Multimodal Emotion Recognition in conversation}

Multimodal emotion recognition in conversations (MERC) aims to infer utterance-level emotions by jointly modeling textual, acoustic, visual, and discourse cues \cite{shou2026comprehensive,shou2022conversational,shou2024adversarial,shou2024low,meng2024deep,shou2025masked,meng2024multi,shou2026graph}. Early approaches primarily rely on feature-level fusion, such as aligning heterogeneous temporal signals before aggregation \cite{yang2017deep,shou2024efficient,shou2025spegcl,ai2026paradigm} or modeling multimodal interactions through tensor-based representations \cite{zadeh2017tensor,shou2025contrastive,shou2025revisiting,shou2025dynamic}.
Subsequent work focuses on improving robustness to incomplete or weak modalities through modality completion, pretrained encoders, decision-level fusion \cite{sun2024similar}, or cross-modal knowledge distillation \cite{yun2024telme}. Related efforts also employ cross-modal alignment or decoupled distillation to strengthen semantic consistency \cite{li2023decoupled,zhang2025multimodal}.

\subsection{Graph Contrastive Learning}

Contrastive learning has emerged as a powerful self-supervised paradigm for graph representation learning, enabling models to capture structural semantics without heavy reliance on annotations \cite{feng2022adversarial,shou2025gsdnet,shou2023graphunet,shou2025graph,shou2025multimodal,shou2025graph}. It works by constructing positive and negative pairs from different views of graph data and training the encoder to pull semantically consistent views together while pushing mismatched ones apart. 
For example, Deep Graph Infomax (DGI) \cite{velivckovic2018deep} maximizes mutual information between local embeddings and a global graph summary. Personalized augmentation strategies \cite{zhang2024graph} adapt perturbations per graph, while GCC \cite{qiu2020gcc} uses ego-networks as structural views to improve topological transferability. Graph-level approaches instead produce whole-graph representations, as in InfoGraph \cite{sun2019infograph}, which aligns global summaries with local patches, and GraphCL \cite{you2020graph}, which systematically evaluates augmentations such as edge perturbation and feature masking. Hybrid strategies like MVGRL \cite{hassani2020contrastive} further combine node-graph and graph-graph comparisons using diffusion-based views.

\section{Method}

\subsection{Overview}
The framework proposed in this paper comprises four components: multimodal feature extraction, dual-space feature decoupling, modality-independent graph learning and modality-specific hypergraph learning, and discourse-level sentiment classification. The overall structure of the framework is shown in Fig. 1. 


\begin{figure*}
    \centering
    \vspace{-3mm}
    \includegraphics[width=1\linewidth]{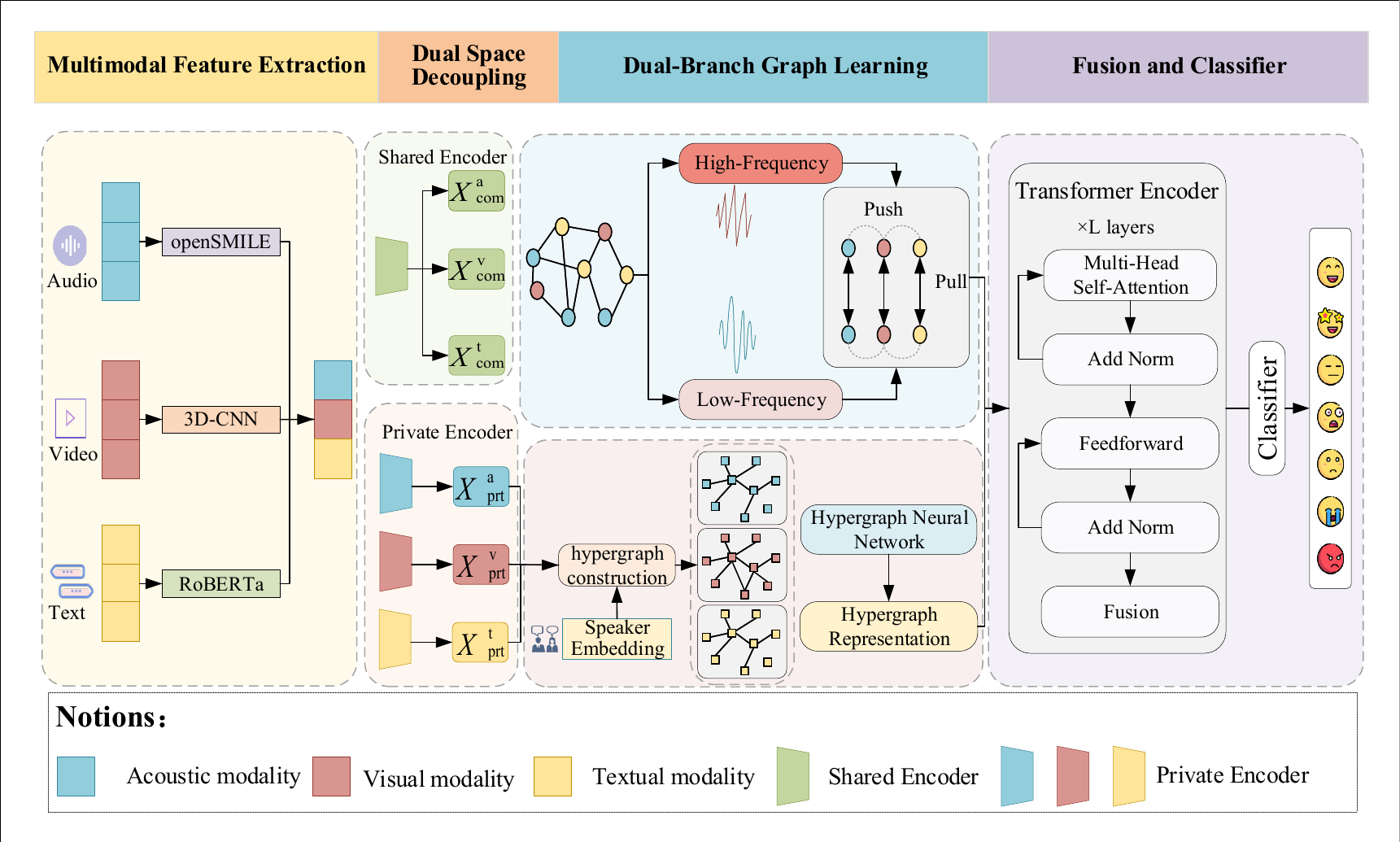}
    \caption{Overall architecture of the proposed framework.}
    \label{fig:frame}
    \vspace{-7mm}
\end{figure*}

\subsection{Multimodal Feature Extraction}
To leverage complementary emotional cues across modalities, we employ modality-specific extractors to encode each utterance into unified representations. A pretrained RoBERTa model is used for textual features, openSMILE descriptors followed by a projection layer for acoustic features, and a 3D-CNN with a projection layer for visual features. Formally, given the $i$-th utterance $u_i$, the corresponding modal features are defined as:

\begin{equation}
\begin{aligned}
\phi_t^i&=\mathrm{RoBERTa}(t_i), \\
\phi_a^i&=\mathrm{Proj}_a(\mathrm{openSMILE}(a_i)), \\
\phi_v^i&=\mathrm{Proj}_v(\mathrm{3D\text{-}CNN}(v_i)),  
\end{aligned}
\end{equation}
where $t_i$, $a_i$, and $v_i$ denote the text, speech signal, and video segment of the $i$-th utterance, respectively. The projection functions $\mathrm{Proj}_a(\cdot)$ and $\mathrm{Proj}_v(\cdot)$ map heterogeneous modal features into a unified latent representation space.

\subsection{Dual-Space Feature Disentanglement}

Multimodal inputs encode both shared information across modalities and modality-specific cues. Directly fusing heterogeneous features can introduce redundancy and weaken discriminative signals. To address this, we propose a dual-space feature disentanglement mechanism that projects each modal feature into a modality-invariant subspace and a modality-specific subspace, enabling separate modeling of shared emotional semantics and modality-specific information.

Specifically, for each modality $m \in \{t,a,v\}$, we employ a cross-modally shared common encoder $E_{\mathrm{com}}(\cdot)$ and a modality-specific private encoder $E_{(m,\mathrm{prt})}(\cdot)$ to map it into a modality-invariant representation $x_{(m,\mathrm{com})}^i$ and a modality-specific representation $x_{(m,\mathrm{prt})}^i$, respectively:

\begin{equation}
x_{(m,\mathrm{com})}^i = E_{\mathrm{com}}(\phi_m^i), \quad x_{(m,\mathrm{prt})}^i = E_{(m,\mathrm{prt})}(\phi_m^i).
\end{equation}

\subsubsection{Reconstruction Loss}

The reconstruction loss is designed to ensure that the concatenation of modality-invariant and modality-specific representations still preserves sufficient information to recover the original input feature. To this end, the shared and private representations are concatenated and fed into a modality-specific decoder $D_m(\cdot)$, and the Euclidean distance between the reconstructed result and the original feature is minimized. The loss is defined as
\begin{equation}
L_{\mathrm{rec}} = \frac{1}{3N} \sum_m \sum_i \left\| \phi_m^i - D_m([x_{(m,\mathrm{com})}^i \parallel x_{(m,\mathrm{prt})}^i]) \right\|_2^2,
\end{equation}
where $D_m(\cdot)$ denotes the modality-specific decoder, and $\parallel$ denotes the feature concatenation operation. 

\subsubsection{Cycle-Consistency Loss}

The reconstruction constraint alone is still insufficient to guarantee that the private representation truly preserves modality-specific information. Therefore, we further introduce a cycle-consistency loss. The key idea is that if the reconstructed feature is passed through the corresponding modality-specific private encoder again, the resulting representation should be as close as possible to the original private representation as

\begin{equation}
\begin{aligned}
L_{\mathrm{cyc}}
=\, \frac{1}{3N}\sum_m\sum_i
\Big\|
x_{(m,\mathrm{prt})}^i  -
E_{(m,\mathrm{prt})}\!\Big(
D_m\!\big(
[x_{(m,\mathrm{com})}^i \parallel x_{(m,\mathrm{prt})}^i]
\big)
\Big)
\Big\|_2^2.
\end{aligned}
\end{equation}

\subsubsection{Margin Loss}

To enhance cross-modal emotional alignment in the modality-invariant subspace, we adopt a triplet-based margin loss. This loss encourages cross-modal representations with the same emotion label to be closer in the shared space, while pushing representations with different emotion labels farther apart, thus improving the discriminability of the shared subspace as
\begin{equation}
\begin{aligned}
L_{\mathrm{mar}}
= \frac{1}{|T|} \sum_{(i,j,k)\in T}
\max \Big(
0,\ &\alpha
- \cos\big(x_{(m(i),\mathrm{com})}^i,\ x_{(m(j),\mathrm{com})}^j\big) \\
&+ \cos\big(x_{(m(i),\mathrm{com})}^i,\ x_{(m(k),\mathrm{com})}^k\big)
\Big).
\end{aligned}
\end{equation}
where $T$ denotes the set of valid triplets, $\alpha$ is the margin hyperparameter, and $\cos(\cdot,\cdot)$ denotes cosine similarity. 

\subsubsection{Orthogonality Loss}

Since the modality-invariant subspace and the modality-specific subspace are responsible for modeling shared information and private information, respectively, the overlap between them should be minimized as much as possible. Based on this consideration, we define an orthogonality loss to constrain the similarity between shared and private representations:
\begin{equation}
L_{\mathrm{ort}} = \frac{1}{3N} \sum_m \sum_i \left| \cos(x_{(m,\mathrm{com})}^i, x_{(m,\mathrm{prt})}^i) \right|.
\end{equation}

\subsubsection{Overall Disentanglement Loss}

By jointly considering the above four constraints, the overall optimization objective of the dual-space feature disentanglement module is defined as
\begin{equation}
L_{\mathrm{dec}} = L_{\mathrm{rec}} + L_{\mathrm{cyc}} + \gamma_1 L_{\mathrm{mar}} + \gamma_2 L_{\mathrm{ort}},
\end{equation}
where $\gamma_1$ and $\gamma_2$ are trade-off coefficients used to balance the influence of different auxiliary loss terms on the disentanglement process. 

\subsection{Modality-Invariant Branch: Fourier Graph Learning}
The modality-invariant representations mainly characterize cross-modal shared emotional semantics. To further model the contextual dependencies among different utterances and modalities in this shared space, we construct a shared interaction graph over the modality-invariant representations and employ a Fourier graph learning module to capture globally consistent patterns and locally complementary emotional dynamics. We first construct the modality-invariant interaction graph $G_{\mathrm{com}}=(V_{\mathrm{com}},E_{\mathrm{com}},A_{\mathrm{com}},X_{\mathrm{com}})$. The node set consists of all modality--utterance pairs, i.e., each utterance is associated with one graph node in the textual, acoustic, and visual modalities, respectively, resulting in $3N$ nodes. The node feature matrix is obtained by stacking the modality-invariant representations of the three modalities $X_{\mathrm{com}}=[X_{(t,\mathrm{com})};X_{(a,\mathrm{com})};X_{(v,\mathrm{com})}]\in\mathbb{R}^{3N\times d}$.

To jointly model intra-modal contextual dependencies and cross-modal semantic correspondence, we introduce two types of edges. The first type connects utterances within the same modality under a sliding window of size $k$, and the second type connects different modalities of the same utterance. Accordingly, the adjacency matrix is defined as

\begin{equation}
A_{\mathrm{com}}[(m-1)N+i,\ (m'-1)N+j]=
\begin{cases}
1, & m=m' \ \text{and}\ |i-j|<k,\\
1, & m\neq m' \ \text{and}\ i=j,\\
0, & \text{otherwise}.
\end{cases}
\end{equation}

A symmetrically normalized adjacency matrix is then obtained as 

\begin{equation}
\tilde{A}_{\mathrm{com}}=D_{\mathrm{com}}^{-1/2}A_{\mathrm{com}}D_{\mathrm{com}}^{-1/2},     
\end{equation}
where $D_{\mathrm{com}}$ is the degree matrix of $A_{\mathrm{com}}$.

Based on the normalized adjacency matrix, the normalized graph Laplacian is defined as

\begin{equation}
L_{\mathrm{com}}=I-\tilde{A}_{\mathrm{com}}.
\end{equation}

Its eigendecomposition is written as

\begin{equation}
L_{\mathrm{com}}=U_{\mathrm{com}}\Lambda_{\mathrm{com}}U_{\mathrm{com}}^{\top}.
\end{equation}

Since conversational emotional dynamics contain both globally smooth trends across turns and locally abrupt or complementary variations, we employ a low-pass filter and a high-pass filter to extract low-frequency and high-frequency components, respectively:

\begin{equation}
X_{\mathrm{com}}^{\ell}=U_{\mathrm{com}}g_{\ell}(\Lambda_{\mathrm{com}})U_{\mathrm{com}}^{\top}X_{\mathrm{com}},
\end{equation}
\begin{equation}
X_{\mathrm{com}}^{h}=U_{\mathrm{com}}g_h(\Lambda_{\mathrm{com}})U_{\mathrm{com}}^{\top}X_{\mathrm{com}},    
\end{equation}
where $g_{\ell}(\cdot)$ and $g_h(\cdot)$ denote the low-pass and high-pass filters, respectively. In this work, we adopt exponential filters, i.e., $g_{\ell}(\lambda)=\exp(-\tau_{\ell}\lambda)$ and $g_h(\lambda)=1-\exp(-\tau_h\lambda)$, where $\tau_{\ell}$ and $\tau_h$ are hyperparameters.

After graph filtering, the outputs are split by modality to obtain modality-specific low-frequency and high-frequency representations. For the $i$-th utterance, the final representation of the shared branch is obtained by concatenating the low-frequency and high-frequency features from all three modalities and then applying a linear transformation:

\begin{equation}
\begin{aligned}
h_{\mathrm{com}}^i
=&\, W_{\mathrm{com}} \big[
x_{(t,\mathrm{com})}^{(\ell,i)} \parallel
x_{(a,\mathrm{com})}^{(\ell,i)} \parallel
x_{(v,\mathrm{com})}^{(\ell,i)} \\
&\qquad \parallel
x_{(t,\mathrm{com})}^{(h,i)} \parallel
x_{(a,\mathrm{com})}^{(h,i)} \parallel
x_{(v,\mathrm{com})}^{(h,i)}
\big]
+ b_{\mathrm{com}},
\end{aligned}
\end{equation}
where $W_{\mathrm{com}}$ and $b_{\mathrm{com}}$ are trainable parameters.

To further improve the discriminability of the shared branch in the frequency domain, we introduce a symmetric InfoNCE contrastive loss between the low-frequency and high-frequency views. Specifically, two projection heads are used to map the low-frequency and high-frequency representations into the contrastive space, denoted by $z_q^{\ell}=P_{\ell}(X_{\mathrm{com}}^{\ell}[q])$ and $z_q^h=P_h(X_{\mathrm{com}}^h[q])$, where $q\in\{1,\ldots,3N\}$ indexes graph nodes. The bidirectional contrastive losses are defined as

\begin{equation}
L_{(\ell\rightarrow h)}=
-\frac{1}{3N}\sum_q
\log
\frac{\exp(\mathrm{sim}(z_q^{\ell},z_q^h)/\tau)}
{\sum_r \exp(\mathrm{sim}(z_q^{\ell},z_r^h)/\tau)},
\end{equation}
\begin{equation}
L_{(h\rightarrow \ell)}=
-\frac{1}{3N}\sum_q
\log
\frac{\exp(\mathrm{sim}(z_q^h,z_q^{\ell})/\tau)}
{\sum_r \exp(\mathrm{sim}(z_q^h,z_r^{\ell})/\tau)},
\end{equation}
where $\mathrm{sim}(\cdot,\cdot)$ denotes cosine similarity and $\tau$ is the temperature parameter. The final contrastive learning objective is
\begin{equation}
L_{\mathrm{cl}}=L_{(\ell\rightarrow h)}+L_{(h\rightarrow \ell)}.
\end{equation}

\subsection{Modality-Specific Branch: Speaker-Aware Hypergraph Learning}

To enhance speaker-related information in the modality-specific representations, the one-hot speaker label $s_i^{\mathrm{onehot}}\in\mathbb{R}^{K}$ is first projected into a dense speaker embedding:
\begin{equation}
e_i^{\mathrm{spk}}=W_{\mathrm{spk}}s_i^{\mathrm{onehot}},
\end{equation}
where $W_{\mathrm{spk}}\in\mathbb{R}^{d\times K}$. 

The speaker embedding is then injected into the private representation to obtain the speaker-aware private feature:
\begin{equation}
\tilde{x}_{(m,\mathrm{prt})}^i=x_{(m,\mathrm{prt})}^i+e_i^{\mathrm{spk}}.
\end{equation}

By stacking all utterances row-wise, we obtain the speaker-aware private feature matrix of each modality, and the overall private feature matrix is formed by stacking the three modalities, i.e., $\tilde{X}_{\mathrm{prt}}=[\tilde{X}_{(t,\mathrm{prt})};\tilde{X}_{(a,\mathrm{prt})};\tilde{X}_{(v,\mathrm{prt})}]\in\mathbb{R}^{3N\times d}$.

Next, we first construct a speaker-aware graph $G_{\mathrm{prt}}=(V_{\mathrm{prt}},E_{\mathrm{prt}},A_{\mathrm{prt}},\tilde{X}_{\mathrm{prt}})$, where $|V_{\mathrm{prt}}|=3N$. Two types of edges are introduced in this graph: intra-speaker temporal edges within a local window, which model the consecutive expression patterns of the same speaker across adjacent turns, and cross-speaker interaction edges between neighboring dialogue turns, which characterize emotional influence and interaction among speakers.

Let the number of edges in $G_{\mathrm{prt}}$ be $M$. We then perform a dual hypergraph transformation to obtain $\mathcal{G}_{\mathrm{prt}}^{*}=(V_{\mathrm{prt}}^{*},E_{\mathrm{prt}}^{*},H_{\mathrm{prt}},X_{\mathrm{prt}}^{*})$. Each edge in the original graph is transformed into a hypergraph vertex, while each node in the original graph is transformed into a hyperedge. For each original graph edge $e=(p,q)$, the corresponding hypergraph vertex feature is defined as

\begin{equation}
x_e^{*}=\frac{\tilde{x}_p+\tilde{x}_q}{2},
\end{equation}
which yields the hypergraph vertex feature matrix $X_{\mathrm{prt}}^{*}\in\mathbb{R}^{M\times d}$. For each original graph node $p$, a hyperedge $\varepsilon_p^{*}$ is constructed to connect all dual vertices adjacent to node $p$, and the corresponding incidence matrix is denoted by $H_{\mathrm{prt}}\in\{0,1\}^{M\times 3N}$.

We further introduce the hyperedge weight matrix $W_e^{*}\in\mathbb{R}^{3N\times 3N}$, the dual-vertex degree matrix $D_v^{*}\in\mathbb{R}^{M\times M}$, and the hyperedge degree matrix $D_e^{*}\in\mathbb{R}^{3N\times 3N}$

\begin{equation}
D_v^{*}(e,e)=\sum_p W_e^{*}(p,p)H_{\mathrm{prt}}(e,p),  
\end{equation}
\begin{equation}
D_e^{*}(p,p)=\sum_e H_{\mathrm{prt}}(e,p).    
\end{equation}

The normalized hypergraph Laplacian is then defined as
\begin{equation}
L_{\mathrm{prt}}^{*}=I_M-(D_v^{*})^{-1/2}H_{\mathrm{prt}}W_e^{*}(D_e^{*})^{-1}H_{\mathrm{prt}}^{\top}(D_v^{*})^{-1/2}.
\end{equation}

To effectively model multi-scale higher-order structures in the spectral domain, we adopt Jacobi polynomials to construct spectral hypergraph filters. Since Jacobi polynomials are defined on $[-1,1]$, the hypergraph Laplacian is first rescaled as

\begin{equation}
\tilde{L}_{\mathrm{prt}}^{*}=\left(\frac{2}{\lambda_{\max}}\right)L_{\mathrm{prt}}^{*}-I_M,
\end{equation}
where $\lambda_{\max}$ is the largest eigenvalue of $L_{\mathrm{prt}}^{*}$. The Jacobi polynomials satisfy $P_0^{(\alpha,\beta)}(x)=1$ and $P_1^{(\alpha,\beta)}(x)=\frac{1}{2}[(\alpha-\beta)+(\alpha+\beta+2)x]$, while higher-order terms follow the standard recurrence relation. Based on this, the $r$-th spectral hypergraph filter is defined as
\begin{equation}
Z_r=P_r^{(\alpha,\beta)}(\tilde{L}_{\mathrm{prt}}^{*})X_{\mathrm{prt}}^{*}W_r,
\end{equation}
where $W_r$ is a learnable parameter matrix.

To effectively aggregate the multi-scale higher-order structural information captured by different filter orders, we further employ an attention mechanism for adaptive fusion. The attention score is computed as
\begin{equation}
e_{(r,e)}=a^{\top}\tanh(W_a Z_r[e]+b_a),
\end{equation}
and the corresponding normalized attention weight is
\begin{equation}
\eta_{(r,e)}=\frac{\exp(e_{(r,e)})}{\sum_q \exp(e_{(q,e)})}.
\end{equation}

The final fused hypergraph representation is then obtained as
\begin{equation}
S_{\mathrm{prt}}^{*}[e]=\sigma\left(\sum_r \eta_{(r,e)} Z_r[e]\right),
\end{equation}
where $S_{\mathrm{prt}}^{*}\in\mathbb{R}^{M\times d}$.

Since $S_{\mathrm{prt}}^{*}$ lies in the dual-vertex space while the downstream classification task requires original modality--utterance node-level representations, we further design a projection-back operation to map the dual hypergraph representation back to the original node space:
\begin{equation}
\bar{S}_{\mathrm{prt}}=(D_e^{*})^{-1}H_{\mathrm{prt}}^{\top}S_{\mathrm{prt}}^{*}\in\mathbb{R}^{3N\times d}.
\end{equation}

The projected representation is then split by modality, and for the $i$-th utterance, the final private representation is obtained by concatenating the three modal representations and applying a linear transformation:
\begin{equation}
h_{\mathrm{prt}}^i=
W_{\mathrm{prt}}
[\bar{s}_{(t,\mathrm{prt})}^i\parallel \bar{s}_{(a,\mathrm{prt})}^i\parallel \bar{s}_{(v,\mathrm{prt})}^i]
+b_{\mathrm{prt}},
\end{equation}
where $\bar{s}_{(m,\mathrm{prt})}^i$ denotes the $i$-th row of the projected representation of modality $m$.

To enhance semantic consistency across multiple dialogue turns of the same speaker, we introduce a speaker-consistency constraint in the private branch. Specifically, for utterance pairs belonging to the same speaker, the distance between their private representations is minimized:
\begin{equation}
L_{(\mathrm{cons,prt})}=
\frac{1}{|\mathcal{P}|}
\sum_{(i,j)\in\mathcal{P}}
\left\|h_{\mathrm{prt}}^i-h_{\mathrm{prt}}^j\right\|_2^2,
\end{equation}
where $\mathcal{P}=\{(i,j)\mid s_i=s_j,\ i<j\}$. It is worth noting that this constraint is not intended to force all utterances from the same speaker to share identical private representations.

Meanwhile, to ensure that the private branch preserves sufficient modality-specific information, we further introduce a private-space reconstruction loss:
\begin{equation}
L_{(\mathrm{rec,prt})}=
\frac{1}{3N}\sum_m\sum_i
\left\|
\tilde{x}_{(m,\mathrm{prt})}^i-
D_{(\mathrm{prt},m)}(\bar{s}_{(m,\mathrm{prt})}^i)
\right\|_2^2,
\end{equation}
where $D_{(\mathrm{prt},m)}(\cdot)$ denotes the modality-specific decoder in the private branch. Accordingly, the overall objective of the private branch is defined as
\begin{equation}
L_{\mathrm{prt}}=L_{(\mathrm{rec,prt})}+\beta L_{(\mathrm{cons,prt})},
\end{equation}
where $\beta$ is a balancing coefficient that controls the strength of the auxiliary speaker-consistency regularization.

\subsection{Transformer-Based Token Fusion and Emotion Classification}
We adopt a transformer-based token fusion module as the final cross-branch aggregation mechanism. We treat the shared representation and the three private representations as separate tokens and allow them to interact through self-attention. For the $i$-th utterance, let $h_i^{com}\in\mathbb{R}^{d_c}$ denote the output of the shared branch, and let $\bar{s}_i^{(t,prt)}$, $\bar{s}_i^{(a,prt)}$, and $\bar{s}_i^{(v,prt)}\in\mathbb{R}^{d_p}$ denote the private representations of the textual, acoustic, and visual branches, respectively. We first project them into a unified fusion space of dimension $d_f$:
\begin{equation}
z_i^{com}=W_{com}^{f}h_i^{com}+b_{com}^{f},
\end{equation}
\begin{equation}
z_i^{m}=W_{m}^{f}\bar{s}_i^{(m,prt)}+b_{m}^{f}, \quad m\in\{t,a,v\}.
\end{equation}

To perform token-level fusion, we prepend a learnable fusion token $z^{fus}\in\mathbb{R}^{d_f}$ and construct the input token sequence as
\begin{equation}
T_i^{(0)}=
\left[
z^{fus};
z_i^{com}+e^{com};
z_i^{t}+e^{t};
z_i^{a}+e^{a};
z_i^{v}+e^{v}
\right]
\in\mathbb{R}^{5\times d_f},
\end{equation}
where $e^{com}$, $e^{t}$, $e^{a}$, and $e^{v}\in\mathbb{R}^{d_f}$ are learnable type embeddings used to distinguish the shared token from the three modality-specific private tokens.

The token sequence is then fed into a transformer encoder to capture cross-branch dependencies. For a self-attention head, the query, key, and value matrices are computed as
\begin{equation}
Q_i=T_i^{(0)}W_Q, \quad K_i=T_i^{(0)}W_K, \quad V_i=T_i^{(0)}W_V,
\end{equation}
where $W_Q$, $W_K$, and $W_V$ are learnable projection matrices. 

The self-attention operation is defined as
\begin{equation}
\mathrm{Attn}(Q_i,K_i,V_i)
=
\mathrm{softmax}\left(\frac{Q_iK_i^{\top}}{\sqrt{d_f}}\right)V_i.
\end{equation}

Using multi-head self-attention, position-wise feed-forward layers, residual connections, and layer normalization, the refined token sequence is obtained as
\begin{equation}
T_i^{(L)}=\mathrm{TransformerEncoder}\left(T_i^{(0)}\right),
\end{equation}
where $L$ denotes the number of transformer layers. We take the output corresponding to the learnable fusion token as the final fused representation:
\begin{equation}
u_i=T_i^{(L)}[0].
\end{equation}

Based on the fused representation, the classifier predicts the emotion distribution as
\begin{equation}
\begin{aligned}
\tilde{u}_i&=\mathrm{ReLU}(W_1u_i+b_1), \\
p_i&=\mathrm{softmax}(W_2\tilde{u}_i+b_2), \\
\hat{y}_i&=\arg\max_{c}p_{i,c},
\end{aligned}
\end{equation}
where $p_i$ is the predicted emotion probability distribution and $\hat{y}_i$ is the predicted emotion label.

The classification objective is defined by the cross-entropy loss:
\begin{equation}
L_{cls}
=
-\frac{1}{N}\sum_{i=1}^{N}\sum_{c}y_{i,c}\log p_{i,c},
\label{eq:47}
\end{equation}
where $y_{i,c}$ denotes the ground-truth label indicator for class $c$.

Finally, the overall training objective is formulated as
\begin{equation}
L_{total}=L_{cls}+\lambda_1L_{dec}+\lambda_2L_{cl}+\lambda_3L_{prt},
\label{eq:48}
\end{equation}
where $L_{dec}$ denotes the decoupling loss, $L_{cl}$ denotes the shared-branch learning objective, $L_{prt}$ denotes the private-branch supervision term, and $\lambda_1$, $\lambda_2$, and $\lambda_3$ are trade-off hyperparameters.

\section{Experiments}
\subsection{Dataset and Evaluation Metrics}
We evaluate the proposed framework on two widely used benchmarks for multimodal emotion recognition in conversations, namely IEMOCAP \cite{busso2008iemocap} and MELD \cite{poria2019meld}. Considering the class-imbalance issue in both datasets, we adopt weighted F1 score (WF1) as the primary evaluation metric. 

\subsection{Baselines}
We compare our proposed model with the following baselines, categorized into three groups: \textit{RNN-based methods}, including DialogueRNN \cite{majumder2019dialoguernn}, HCL \cite{yang2022hybrid}, and MVN \cite{ma2022multi}; \textit{graph-based methods}, including MMGCN \cite{hu2021mmgcn}, MMDFN \cite{hu2022mm}, GA2MIF \cite{li2023ga2mif}; and \textit{graph transformer-based methods}, including CMCF-SRNet \cite{zhang2023cross}.

\section{Experimental Results}
\subsection{Quantitative Results}
Table~\ref{tab:results} reports the performance comparison of our method against several strong baselines on the IEMOCAP and MELD datasets in terms of per-class F1-scores and overall weighted F1-score (WF1). On IEMOCAP, the proposed model achieves a WF1 score of 70.81, outperforming all compared baselines. At the class level, our method achieves the best F1 scores on \textit{happy} and \textit{angry}, while also remaining competitive on the other emotion categories.  On MELD, the proposed method achieves the best WF1 scores on all reported emotion categories. Since MELD contains more complex multi-party conversational interactions and stronger speaker interference than IEMOCAP, these results further verify the robustness of our framework in challenging conversational scenarios.

\begin{table*}[!t]
\centering
\vspace{-2mm}
\caption{Performance (F1 score) comparison of different methods on the IEMOCAP and MELD datasets. The best result in each column is in bold.}
\resizebox{0.95\textwidth}{!}{
\begin{tabular}{l|ccccccc|cccccc}
\hline
\multirow{2}{*}{Methods} & \multicolumn{6}{c}{IEMOCAP} & \multicolumn{6}{c}{MELD} &  \\
\cline{2-14}
& Happy & Sad & Neutral & Angry & Excited & Frustrated & WF1  
& Neutral & Surprise & Sadness & Happy & Anger & WF1 \\
\hline \hline
DialogueRNN     & 32.20 & 80.26 & 57.89 & 62.82 & 73.87 & 59.76 & {62.89} & 76.97 & 47.69 & 20.41 & 50.92 & 45.52 & 57.66 \\
HCL             & 48.97 & 82.21 & 60.88 & 66.72 & 69.43 & \textbf{68.73} & {68.73} & - & - & - & - & - & {63.89} \\
MVN             & 55.75 & 73.30 & 61.88 & 65.96 & 69.50 & 64.21 & 65.44 &  76.65 & 53.18 & 21.82 & 53.62 & 42.55 & 59.03 \\
SCMM            & 45.37 & 78.76 & 63.54 & 66.05 & \textbf{76.70} & 66.18 & 67.53 & - & - & - & - & - & 59.44 \\
\hline
MMGCN           & 45.14 & 77.16 & 64.36 & 68.82 & 74.71 & 61.40 & 66.26 & 76.33 & 48.15 & 26.74 & 53.02 & 46.09 & 58.31 \\
MMDFN           & 42.22 & 78.98 & 66.42 & 69.77 & 75.56 & 66.33 & 68.18 & 77.76 & 50.69 & 22.93 & 54.78 & 42.87 & 59.46 \\
GA2MIF          & 46.15 & \textbf{84.50} & 68.08 & 70.29 & 75.99 & 66.49 & 70.00 & 77.62 & 49.08 & 27.18 & 51.87 & 45.62 & 60.83 \\
CMCF-SRNet      & 52.20 & 80.90 & \textbf{68.80} & {70.30} & \textbf{76.70} & 61.60 & 69.60 & 77.20 & 52.90 & 36.00 & 55.80 & 43.90 & 61.80 \\
\hline
COGMEN          & 51.91 & 81.72 & 68.61 & 66.02 & 75.31 & 58.23 & 67.03 & 75.31 & 46.75 & 33.52 & 54.98 & 45.81 & 58.66 \\
Ours & \textbf{60.84} &	80.71 & 67.16 & \textbf{71.43} & 74.69 & 68.60 & \textbf{70.81} & \textbf{79.04} & \textbf{60.10} & \textbf{39.65} & \textbf{62.83} & \textbf{53.87} & \textbf{65.70}\\
\hline
\end{tabular}
}
\vspace{-2mm}
\label{tab:results}
\end{table*}

\begin{table}[htbp]  
    \centering
    \small
    \caption{Ablation studies on IEMOCAP and MELD datasets.}  
    \label{tab:ablation}
    \renewcommand{\arraystretch}{1}  
    \resizebox{0.95\textwidth}{!}{  
    \begin{tabular}{l c c c c} 
    \hline
    \multirow{2}{*}{Methods} & \multicolumn{2}{c}{IEMOCAP} & \multicolumn{2}{c}{MELD} \\ 
    \cmidrule{2-5} 
    & \textbf{Acc} & \textbf{WF1} & \textbf{Acc} & \textbf{WF1} \\ 
    \hline
    Ours           & 70.63 & 70.81 & 66.48 & 65.70 \\ 
    \hline
    - w/o Decoupler      & 67.16 & 67.38 & 63.14 & 62.41 \\
    - w/o Shared Branch (FourierGraph)    & 65.83 & 66.01 & 61.87 & 61.08 \\
    - w/o Private Branch (HGNN)     & 68.02 & 68.19 & 64.23 & 63.47 \\ 
    - w/o Transformer Fusion   & 68.91 & 69.07 & 64.89 & 64.12 \\ 
    \hline
    \end{tabular}
    }
    \vspace{-4mm}
\end{table}

\subsection{Ablation Study}
We conduct ablation studies on IEMOCAP and MELD to evaluate the contribution of each component. As shown in Table~\ref{tab:ablation}, the full model achieves the best performance on both datasets, validating the overall design. Removing the decoupling module leads to clear performance drops, indicating that separating modality-invariant and modality-specific information helps reduce redundancy and improve representation quality. Excluding the shared branch results in the largest degradation, highlighting its critical role in capturing cross-modal semantics and global interaction patterns. Removing the private branch also degrades performance, with WF1 dropping to 68.19 on IEMOCAP and 63.47 on MELD, demonstrating the importance of modality-specific relational modeling. Additionally, removing the transformer fusion module consistently lowers performance on both datasets. Although the decrease is smaller, it still shows that transformer-based fusion is more effective in integrating shared and modality-specific representations.

\section{Conclusions}
In this paper, we address the limitations of existing methods in modeling long-range dependencies and high-order interactions for multimodal conversational emotion recognition by proposing a unified framework that integrates dual-space feature disentanglement with dual-branch graph learning. Specifically, a shared encoder and modality-specific encoders are used to separate modality-invariant emotional semantics from modality-specific discriminative cues. Based on these representations, a shared interaction graph and a modality-specific hypergraph branch are constructed. A Fourier graph learning module captures global consistency and long-range dependencies, while an HGNN-based branch models modality-dependent and high-order speaker interactions. Finally, a transformer-based fusion module dynamically integrates the shared and specific representations for utterance-level emotion classification. Experiments on the IEMOCAP and MELD datasets demonstrate that the proposed framework achieves strong and robust performance, validating the effectiveness of disentangled representation learning and dual-branch relational modeling.

\bibliographystyle{spmpsci_unsrt}
\bibliography{refx}

\begin{thebibliography}{10}
\providecommand{\url}[1]{{#1}}
\providecommand{\urlprefix}{URL }
\expandafter\ifx\csname urlstyle\endcsname\relax
  \providecommand{\doi}[1]{DOI~\discretionary{}{}{}#1}\else
  \providecommand{\doi}{DOI~\discretionary{}{}{}\begingroup
  \urlstyle{rm}\Url}\fi

\bibitem{zhang2025multimodal}
Zhang, S., Liu, J., Jiao, Y., Zhang, Y., Chen, L., Li, K.: A multimodal
  semantic fusion network with cross-modal alignment for multimodal sentiment
  analysis.
\newblock ACM Transactions on Multimedia Computing, Communications and
  Applications  (2025)

\bibitem{shou2025cilf}
Shou, Y., Meng, T., Ai, W., Yin, N., Li, K.: Cilf-ciae: Clip-driven
  image--language fusion for correcting inverse age estimation.
\newblock Neural Networks p. 108518 (2025)

\bibitem{shou2026dual}
Shou, Y., Zhou, J., Meng, T., Ai, W., Li, K.: Dual-branch graph domain
  adaptation for cross-scenario multi-modal emotion recognition.
\newblock arXiv preprint arXiv:2603.26840  (2026)

\bibitem{shou2026comprehensive}
Shou, Y., Meng, T., Ai, W., Fu, F., Yin, N., Li, K.: A comprehensive survey on
  multi-modal conversational emotion recognition with deep learning.
\newblock ACM Transactions on Information Systems \textbf{44}(2), 1--48 (2026)

\bibitem{shou2022conversational}
Shou, Y., Meng, T., Ai, W., Yang, S., Li, K.: Conversational emotion
  recognition studies based on graph convolutional neural networks and a
  dependent syntactic analysis.
\newblock Neurocomputing \textbf{501}, 629--639 (2022)

\bibitem{shou2024adversarial}
Shou, Y., Meng, T., Ai, W., Zhang, F., Yin, N., Li, K.: Adversarial alignment
  and graph fusion via information bottleneck for multimodal emotion
  recognition in conversations.
\newblock Information Fusion \textbf{112}, 102,590 (2024)

\bibitem{shou2024low}
Shou, Y., Liu, H., Cao, X., Meng, D., Dong, B.: A low-rank matching attention
  based cross-modal feature fusion method for conversational emotion
  recognition.
\newblock IEEE Transactions on Affective Computing \textbf{16}(2), 1177--1189
  (2024)

\bibitem{meng2024deep}
Meng, T., Shou, Y., Ai, W., Yin, N., Li, K.: Deep imbalanced learning for
  multimodal emotion recognition in conversations.
\newblock IEEE Transactions on Artificial Intelligence \textbf{5}(12),
  6472--6487 (2024)

\bibitem{shou2025masked}
Shou, Y., Cao, X., Liu, H., Meng, D.: Masked contrastive graph representation
  learning for age estimation.
\newblock Pattern Recognition \textbf{158}, 110,974 (2025)

\bibitem{meng2024multi}
Meng, T., Shou, Y., Ai, W., Du, J., Liu, H., Li, K.: A multi-message passing
  framework based on heterogeneous graphs in conversational emotion
  recognition.
\newblock Neurocomputing \textbf{569}, 127,109 (2024)

\bibitem{shou2026graph}
Shou, Y., Ai, W., Meng, T., Li, K.: Graph diffusion models: A comprehensive
  survey of methods and applications.
\newblock Computer Science Review \textbf{59}, 100,854 (2026)

\bibitem{yang2017deep}
Yang, X., Ramesh, P., Chitta, R., Madhvanath, S., Bernal, E.A., Luo, J.: Deep
  multimodal representation learning from temporal data.
\newblock In: Proceedings of the IEEE conference on computer vision and pattern
  recognition, pp. 5447--5455 (2017)

\bibitem{shou2024efficient}
Shou, Y., Ai, W., Du, J., Meng, T., Liu, H., Yin, N.: Efficient long-distance
  latent relation-aware graph neural network for multi-modal emotion
  recognition in conversations.
\newblock arXiv preprint arXiv:2407.00119  (2024)

\bibitem{shou2025spegcl}
Shou, Y., Cao, X., Meng, D.: Spegcl: Self-supervised graph spectrum contrastive
  learning without positive samples.
\newblock IEEE Transactions on Neural Networks and Learning Systems  (2025)

\bibitem{ai2026paradigm}
Ai, W., Tan, Y., Shou, Y., Meng, T., Chen, H., He, Z., Li, K.: The paradigm
  shift: A comprehensive survey on large vision language models for multimodal
  fake news detection.
\newblock Computer Science Review \textbf{60}, 100,893 (2026)

\bibitem{zadeh2017tensor}
Zadeh, A., Chen, M., Poria, S., Cambria, E., Morency, L.P.: Tensor fusion
  network for multimodal sentiment analysis.
\newblock arXiv preprint arXiv:1707.07250  (2017)

\bibitem{shou2025contrastive}
Shou, Y., Lan, H., Cao, X.: Contrastive graph representation learning with
  adversarial cross-view reconstruction and information bottleneck.
\newblock Neural Networks \textbf{184}, 107,094 (2025)

\bibitem{shou2025revisiting}
Shou, Y., Meng, T., Ai, W., Li, K.: Revisiting multi-modal emotion learning
  with broad state space models and probability-guidance fusion.
\newblock In: Joint European Conference on Machine Learning and Knowledge
  Discovery in Databases, pp. 509--525. Springer (2025)

\bibitem{shou2025dynamic}
Shou, Y., Meng, T., Ai, W., Li, K.: Dynamic graph neural ode network for
  multi-modal emotion recognition in conversation.
\newblock In: Proceedings of the 31st International Conference on Computational
  Linguistics, pp. 256--268 (2025)

\bibitem{sun2024similar}
Sun, Y., Liu, Z., Sheng, Q.Z., Chu, D., Yu, J., Sun, H.: Similar modality
  completion-based multimodal sentiment analysis under uncertain missing
  modalities.
\newblock Information Fusion \textbf{110}, 102,454 (2024)

\bibitem{yun2024telme}
Yun, T., Lim, H., Lee, J., Song, M.: Telme: Teacher-leading multimodal fusion
  network for emotion recognition in conversation.
\newblock arXiv preprint arXiv:2401.12987  (2024)

\bibitem{li2023decoupled}
Li, Y., Wang, Y., Cui, Z.: Decoupled multimodal distilling for emotion
  recognition.
\newblock In: Proceedings of the IEEE/CVF conference on computer vision and
  pattern recognition, pp. 6631--6640 (2023)

\bibitem{feng2022adversarial}
Feng, S., Jing, B., Zhu, Y., Tong, H.: Adversarial graph contrastive learning
  with information regularization.
\newblock In: Proceedings of the ACM web conference 2022, pp. 1362--1371 (2022)

\bibitem{shou2025gsdnet}
Shou, Y., Yao, J., Meng, T., Ai, W., Chen, C., Li, K.: Gsdnet: Revisiting
  incomplete multimodality-diffusion emotion recognition from the perspective
  of graph spectrum.
\newblock In: Proceedings of the Thirty-Fourth International Joint Conference
  on Artificial Intelligence, IJCAI-25. International Joint Conferences on
  Artificial Intelligence Organization, pp. 6182--6190 (2025)

\bibitem{shou2023graphunet}
Shou, Y., Ai, W., Meng, T., Zhang, F., Li, K.: Graphunet: Graph make strong
  encoders for remote sensing segmentation.
\newblock In: 2023 IEEE 29th International Conference on Parallel and
  Distributed Systems (ICPADS), pp. 2734--2737. IEEE (2023)

\bibitem{shou2025graph}
Shou, Y., Cao, X., Yan, P., Hui, Q., Zhao, Q., Meng, D.: Graph domain
  adaptation with dual-branch encoder and two-level alignment for whole slide
  image-based survival prediction.
\newblock In: Proceedings of the IEEE/CVF International Conference on Computer
  Vision, pp. 19,925--19,935 (2025)

\bibitem{shou2025multimodal}
Shou, Y., Meng, T., Ai, W., Li, K.: Multimodal large language models meet
  multimodal emotion recognition and reasoning: A survey.
\newblock arXiv preprint arXiv:2509.24322  (2025)

\bibitem{velivckovic2018deep}
Veli{\v{c}}kovi{\'c}, P., Fedus, W., Hamilton, W.L., Li{\`o}, P., Bengio, Y.,
  Hjelm, R.D.: Deep graph infomax.
\newblock arXiv preprint arXiv:1809.10341  (2018)

\bibitem{zhang2024graph}
Zhang, X., Tan, Q., Huang, X., Li, B.: Graph contrastive learning with
  personalized augmentation.
\newblock IEEE Transactions on Knowledge and Data Engineering \textbf{36}(11),
  6305--6316 (2024)

\bibitem{qiu2020gcc}
Qiu, J., Chen, Q., Dong, Y., Zhang, J., Yang, H., Ding, M., Wang, K., Tang, J.:
  Gcc: Graph contrastive coding for graph neural network pre-training.
\newblock In: Proceedings of the 26th ACM SIGKDD international conference on
  knowledge discovery \& data mining, pp. 1150--1160 (2020)

\bibitem{sun2019infograph}
Sun, F.Y., Hoffmann, J., Verma, V., Tang, J.: Infograph: Unsupervised and
  semi-supervised graph-level representation learning via mutual information
  maximization.
\newblock arXiv preprint arXiv:1908.01000  (2019)

\bibitem{you2020graph}
You, Y., Chen, T., Sui, Y., Chen, T., Wang, Z., Shen, Y.: Graph contrastive
  learning with augmentations.
\newblock Advances in neural information processing systems \textbf{33},
  5812--5823 (2020)

\bibitem{hassani2020contrastive}
Hassani, K., Khasahmadi, A.H.: Contrastive multi-view representation learning
  on graphs.
\newblock In: International conference on machine learning, pp. 4116--4126.
  PMLR (2020)

\bibitem{busso2008iemocap}
Busso, C., Bulut, M., Lee, C.C., Kazemzadeh, A., Mower, E., Kim, S., Chang,
  J.N., Lee, S., Narayanan, S.S.: Iemocap: Interactive emotional dyadic motion
  capture database.
\newblock Language resources and evaluation \textbf{42}, 335--359 (2008)

\bibitem{poria2019meld}
Poria, S., Hazarika, D., Majumder, N., Naik, G., Cambria, E., Mihalcea, R.:
  Meld: A multimodal multi-party dataset for emotion recognition in
  conversations.
\newblock In: Proceedings of the 57th Annual Meeting of the Association for
  Computational Linguistics. ACL (2019)

\bibitem{majumder2019dialoguernn}
Majumder, N., Poria, S., Hazarika, D., Mihalcea, R., Gelbukh, A., Cambria, E.:
  Dialoguernn: An attentive rnn for emotion detection in conversations.
\newblock In: Proceedings of the AAAI conference on artificial intelligence,
  vol.~33, pp. 6818--6825 (2019)

\bibitem{yang2022hybrid}
Yang, L., Shen, Y., Mao, Y., Cai, L.: Hybrid curriculum learning for emotion
  recognition in conversation.
\newblock In: Proceedings of the AAAI conference on artificial intelligence,
  vol.~36, pp. 11,595--11,603 (2022)

\bibitem{ma2022multi}
Ma, H., Wang, J., Lin, H., Pan, X., Zhang, Y., Yang, Z.: A multi-view network
  for real-time emotion recognition in conversations.
\newblock Knowledge-Based Systems \textbf{236}, 107,751 (2022)

\bibitem{hu2021mmgcn}
Hu, J., Liu, Y., Zhao, J., Jin, Q.: Mmgcn: Multimodal fusion via deep graph
  convolution network for emotion recognition in conversation.
\newblock arXiv preprint arXiv:2107.06779  (2021)

\bibitem{hu2022mm}
Hu, D., Hou, X., Wei, L., Jiang, L., Mo, Y.: Mm-dfn: Multimodal dynamic fusion
  network for emotion recognition in conversations.
\newblock In: ICASSP 2022-2022 IEEE International Conference on Acoustics,
  Speech and Signal Processing (ICASSP), pp. 7037--7041. IEEE (2022)

\bibitem{li2023ga2mif}
Li, J., Wang, X., Lv, G., Zeng, Z.: Ga2mif: Graph and attention based two-stage
  multi-source information fusion for conversational emotion detection.
\newblock IEEE Transactions on affective computing \textbf{15}(1), 130--143
  (2023)

\bibitem{zhang2023cross}
Zhang, X., Li, Y.: A cross-modality context fusion and semantic refinement
  network for emotion recognition in conversation.
\newblock In: Proceedings of the 61st Annual Meeting of the Association for
  Computational Linguistics (Volume 1: Long Papers), pp. 13,099--13,110 (2023)

\end{thebibliography}

\end{document}